\input harvmac
\input amssym

\lref\AcharyaGY{
  B.~Acharya and E.~Witten,
  ``Chiral fermions from manifolds of G(2) holonomy,''
  arXiv:hep-th/0109152.
}

\lref\ArkaniHamedDZ{
  N.~Arkani-Hamed, L.~Motl, A.~Nicolis and C.~Vafa,
  ``The string landscape, black holes and gravity as the weakest force,''
  arXiv:hep-th/0601001.
}

\lref\ChoiXN{
  K.~Choi,
  ``String or M theory axion as a quintessence,''
  Phys.\ Rev.\ D {\bf 62}, 043509 (2000)
  [arXiv:hep-ph/9902292].
}

\lref\WenJZ{
  X.~G.~Wen and E.~Witten,
  ``World Sheet Instantons And The Peccei-Quinn Symmetry,''
  Phys.\ Lett.\ B {\bf 166}, 397 (1986).
}

\lref\DineBQ{ M.~Dine, N.~Seiberg, X.~G.~Wen and E.~Witten,
``Nonperturbative Effects On The String World Sheet. 2,'' Nucl.\
Phys.\ B {\bf 289}, 319 (1987).
}

\lref\DineZY{ M.~Dine, N.~Seiberg, X.~G.~Wen and E.~Witten,
``Nonperturbative Effects On The String World Sheet,'' Nucl.\ Phys.\
B {\bf 278}, 769 (1986).
}

\lref\EastherZR{
  R.~Easther and L.~McAllister,
  ``Random matrices and the spectrum of N-flation,''
  JCAP {\bf 0605}, 018 (2006)
  [arXiv:hep-th/0512102].
}

\lref\BanksSG{
  T.~Banks and M.~Dine,
  ``Coping with strongly coupled string theory,''
  Phys.\ Rev.\ D {\bf 50}, 7454 (1994)
  [arXiv:hep-th/9406132].}

\lref\banksone{T. Banks and M. Dine, "Couplings and Scales in
Strongly Coupled Heterotic String Theory," Nucl. Phys. {\bf B479}
(1996) 173, arXiv:hep-th/9605136.}

\lref\bankstwo{T. Banks and M. Dine, "The Cosmology of String
Theoretic Axions," Nucl. Phys. {\bf B505} (1997) 445,
arXiv:hep-th/9608197.}

\lref\bankscosmo{T. Banks, M. Dine and M. Graesser, "Supersymmetry,
Axions and Cosmology," Phys. Rev. {\bf D68} (2003) 0705011,
arXiv:hep-ph/0210256.}

\lref\bankslarge{T. Banks, M. Dine, P. J. Fox and E. Gorbatov, "On
the Possibility of Large Axion Decay Constants," JCAP {\bf 0306}
(2003) 001, arXiv:hep-th/0303252.}

\lref\CarrollZI{
  S.~M.~Carroll,
  ``Quintessence and the rest of the world,''
  Phys.\ Rev.\ Lett.\  {\bf 81}, 3067 (1998)
  [arXiv:astro-ph/9806099].
}

\lref\ChoiXN{ K.~Choi, ``String or M theory axion as a
quintessence,'' Phys.\ Rev.\ D {\bf 62}, 043509 (2000)
[arXiv:hep-ph/9902292].
}

\lref\HallXB{
  L.~J.~Hall, Y.~Nomura and S.~J.~Oliver,
  ``Evolving dark energy with w not equal -1,''
  Phys.\ Rev.\ Lett.\  {\bf 95}, 141302 (2005)
  [arXiv:astro-ph/0503706].
}

\lref\ItzhakiRE{
  N.~Itzhaki,
  ``A comment on technical naturalness and the cosmological constant,''
  arXiv:hep-th/0604190.
}

\lref\KachruAW{
  S.~Kachru, R.~Kallosh, A.~Linde and S.~P.~Trivedi,
  `De Sitter vacua in string theory,''
  Phys.\ Rev.\ D {\bf 68}, 046005 (2003)
  [arXiv:hep-th/0301240].
}

\lref\PolchinskiRR{
  J.~Polchinski,
  ``String theory. Vol. 2: Superstring theory and beyond,''
}

\lref\SteinhardtBF{
  P.~J.~Steinhardt and N.~Turok,
  ``Why the cosmological constant is small and positive,''
  arXiv:astro-ph/0605173.
}

\lref\GiddingsYU{
  S.~B.~Giddings, S.~Kachru and J.~Polchinski,
  ``Hierarchies from fluxes in string compactifications,''
  Phys.\ Rev.\ D {\bf 66}, 106006 (2002)
  [arXiv:hep-th/0105097].
}
\lref\BoussoXA{
  R.~Bousso and J.~Polchinski,
  ``Quantization of four-form fluxes and dynamical neutralization of the
  cosmological constant,''
  JHEP {\bf 0006}, 006 (2000)
  [arXiv:hep-th/0004134].
}

\lref\ConlonKI{
  J.~P.~Conlon, F.~Quevedo and K.~Suruliz,
  ``Large-volume flux compactifications: Moduli spectrum and D3/D7 soft
  supersymmetry breaking,''
  JHEP {\bf 0508}, 007 (2005)
  [arXiv:hep-th/0505076].
}

\lref\ConlonTQ{
  J.~P.~Conlon,
  ``The QCD axion and moduli stabilisation,''
  arXiv:hep-th/0602233.
}

\nref\bb{L. F. Abbott and P. Sikivie, ``A Cosmological Bound On The
Invisible Axion,'' Phys. Lett. {\bf B120}
(1983) 133.}%
\nref\cc{M. Dine and W. Fischler, ``The Not So Harmless Axion,''
 Phys. Lett. {\bf B120} (1983) 137.}%

\lref\KachruAW{
  S.~Kachru, R.~Kallosh, A.~Linde and S.~P.~Trivedi,
  ``De Sitter vacua in string theory,''
  Phys.\ Rev.\ D {\bf 68}, 046005 (2003)
  [arXiv:hep-th/0301240].
}
\lref\mcallister{
  R.~Easther and L.~McAllister,
  ``Random matrices and the spectrum of N-flation,''
  arXiv:hep-th/0512102.
}

\lref\FriedmannTY{
  T.~Friedmann and E.~Witten,
  ``Unification scale, proton decay, and manifolds of G(2) holonomy,''
  Adv.\ Theor.\ Math.\ Phys.\  {\bf 7}, 577 (2003)
  [arXiv:hep-th/0211269].
}

\lref\PreskillCY{
  J.~Preskill, M.~B.~Wise and F.~Wilczek,
  ``Cosmology Of The Invisible Axion,''
  Phys.\ Lett.\ B {\bf 120}, 127 (1983).
}

\lref\SusskindKW{
  L.~Susskind,
  ``The anthropic landscape of string theory,''
  arXiv:hep-th/0302219.
}
\lref\KaloperAJ{
  N.~Kaloper and L.~Sorbo,
  ``Of pNGB QuiNtessence,''
  arXiv:astro-ph/0511543.
}

\lref\KoldaWQ{
  C.~F.~Kolda and D.~H.~Lyth,
  ``Quintessential difficulties,''
  Phys.\ Lett.\ B {\bf 458}, 197 (1999)
  [arXiv:hep-ph/9811375].
}
\lref\DimopoulosAC{
  S.~Dimopoulos, S.~Kachru, J.~McGreevy and J.~G.~Wacker,
  ``N-flation,''
  arXiv:hep-th/0507205.
}

\lref\DeWolfeUU{
  O.~DeWolfe, A.~Giryavets, S.~Kachru and W.~Taylor,
  JHEP {\bf 0507}, 066 (2005)
  [arXiv:hep-th/0505160].
}

\lref\SteinhardtBF{
  P.~J.~Steinhardt and N.~Turok,
  ``Why the cosmological constant is small and positive,''
  arXiv:astro-ph/0605173.
}

\lref\VafaUI{
  C.~Vafa,
  ``The string landscape and the swampland,''
  arXiv:hep-th/0509212.
}

\lref\WeinbergDV{
  S.~Weinberg,
  ``Anthropic Bound On The Cosmological Constant,''
  Phys.\ Rev.\ Lett.\  {\bf 59}, 2607 (1987).
}

\lref\wittensc{E. Witten, ``Strong Coupling Expansion Of Calabi-Yau
Compactification,'' Nucl. Phys. {\bf B471} (1996) 135,
hep-th/9602070.}

\lref\axionpaper{P. Svr\v{c}ek and E. Witten, ``Axions in String
Theory,'' hep-th/0605206.}

\lref\lindechaotic{A. D. Linde, ``Chaotic Inflation,'' Phys. Lett.
{\bf B129} (1983) 177.}

\lref\WangFA{
  L.~M.~Wang, R.~R.~Caldwell, J.~P.~Ostriker and P.~J.~Steinhardt,
  ``Cosmic Concordance and Quintessence,''
  Astrophys.\ J.\  {\bf 530}, 17 (2000)
  [arXiv:astro-ph/9901388].
}

\lref\WittenZK{
  E.~Witten,
  ``The cosmological constant from the viewpoint of string theory,''
  arXiv:hep-ph/0002297.
}

\lref\ZlatevTR{
  I.~Zlatev, L.~M.~Wang and P.~J.~Steinhardt,
  ``Quintessence, Cosmic Coincidence, and the Cosmological Constant,''
  Phys.\ Rev.\ Lett.\  {\bf 82}, 896 (1999)
  [arXiv:astro-ph/9807002].
}

\def\GeV{{\rm GeV}}
\def\CN{{\cal N}}

\Title{hep-th/0607086} {\vbox{\centerline{}
\bigskip
\centerline{Cosmological Constant and Axions in String Theory }}}
\smallskip\centerline{Peter Svr\v{c}ek}
\smallskip\centerline{\it Department of Physics and SLAC, Stanford
University}\centerline{\it Stanford CA 94305/94309 USA}

\medskip

String theory axions appear to be promising candidates for
explaining cosmological constant via quintessence. In this paper, we
study conditions on the string compactifications under which axion
quintessence can happen. For sufficiently large number of axions,
cosmological constant can be accounted for as the potential energy
of axions that have not yet relaxed to their minima. In
compactifications that incorporate unified models of particle
physics, the height of the axion potential can naturally fall close
to the observed value of cosmological constant.

\noindent \Date{July, 2006}

\newsec{Introduction}

Ever since the observation of nonzero cosmological constant
\eqn\vacpe{\Lambda \sim e^{-280}M_P^4,} where $M_P=1/\sqrt{8\pi
G_N}=2.4\times 10^{18}\GeV,$ it has been a puzzle why the
cosmological constant is so small compared to the Planck scale and
yet nonzero. Theoretically, one would expect the cosmological
constant to be around the Planck scale.  It has been argued that
even if we were able to set the cosmological constant to its present
value at low orders in perturbation theory, higher order radiative
corrections generate cosmological constant of the order of the
Planck scale $\delta \Lambda\sim M_P^4.$  This would give
cosmological constant which is $120$ orders of magnitude larger than
the observed value.

In one approach to this problem, the cosmological constant is
attributed to the vacuum energy of a scalar field, called
quintessence, that has not yet relaxed to its vacuum
\refs{\ZlatevTR, \WangFA}. The vacuum is assumed to have zero or
vanishingly small cosmological constant due to an unknown physical
principle. To account for the cosmological constant this way, it is
necessary to explain why the scalar potential takes values close to
the measured cosmological constant and why the scalar field is
lighter than the Hubble scale $H\sim e^{-140} M_P$, so that it
contributes to dark energy rather than dark matter.

Such light scalars are expected to have Planck suppressed couplings
to matter fields that would lead to observable long-range forces and
to variation of constants of nature \CarrollZI.\  Stringent tests of
equivalence principle and of variation of constants of nature
constrain the scalar couplings to matter to anomalously small
values, imposing fine-tuning on the quintessence models.
Furthermore, even if these bounds could be evaded, additional
fine-tuning is necessary  since radiative corrections are expected
to spoil the flatness of the scalar potential and to generate large
mass for the scalar field \KoldaWQ.\

To evade these problems, it has been suggested to consider
pseudoscalar fields called axions. These are protected by global
symmetry $a\rightarrow a+c$ where $a$ is the axion and $c$ is an
arbitrary constant. The shift symmetry suppresses the couplings of
axion to matter which relaxes the above observational constraints.
Effects that break the shift symmetry generate axion mass and
potential. In string theory, the shift symmetry is broken only on
nonperturbative level. The nonperturbative effects that break the
shift symmetry are exponentially suppressed and can naturally lead
to potentials many orders of magnitude below Planckian energy
densities.

In string compactifications that incorporate unified models of
particle physics, matching to the observed value of gauge couplings
and the unification scale leads to further constraints. These fix
some parameters of the compactification and hence allow us to
estimate the scale of the instanton generated axion potential. We
perform this estimate and find that in some string
compactifications, such as of heterotic M-theory, the potential
comes out roughly in the range preferred for quintessence. In
others, such as weakly coupled heterotic string, the potential comes
out too large, hence disfavoring axion quintessence. Thus axion
quintessence may relate the hierarchy between cosmological constant
and the Planck scale to the hierarchy between the unification scale
and the Planck scale and can differentiate between different
compactifications.

In summary, string theory axions can both be very light so that they
contribute to dark energy and can evade the observational bounds
coming from long range forces and the variation of constants of
nature \WittenZK.\ Besides getting the correct value of cosmological
constant, one also has to make sure that the axion is slow rolling
down its potential so that it acts like dark energy rather than like
dark matter. As we will discuss below, this puts further constraints
on parameters of the axion in addition to the above requirement that
the axion has potential energy comparable to the cosmological
constant.

\nref\mcgreevy{J. McGreevy, S. Kachru and P. Svr\v{c}ek, ``Bounds on
Masses of Bulk Fields in String Compactifications,''
hep-th/0601111.}

In this paper, we study whether these conditions are met by string
theory axions. We find that characteristics of string theory axions
are uniform across different compactifications. In the case of one
axion, the axion coupling parameter $F_a$ has to be comparable or
larger than the Planck mass $M_P.$ We will argue that in string
theory, there is an upper bound on $F_a$, which for very light
axions restricts $F_a$ below the Planck scale. Hence, in string
theory a single axion quintessence does not seem to be natural.
Quintessence could be achieved only by fine-tuning the initial
conditions of the axion to make it sit near the top of its potential
for a long time to simulate cosmological constant. This gives an
illustration of a situation, in which a mechanism that is consistent
as an effective field theory does not seem to work in string theory
\refs{\VafaUI, \ArkaniHamedDZ, \mcgreevy}.

Although single axion quintessence seems to be excluded in string
theory, quintessence with many axions is possible in some string
compactifications. Whether this happens depends on the axion
parameters and on the number of axions that one gets in a string
compactification. In the following, we will determine these
conditions. We find that with at least roughly $10^{4-5}$ axions,
the quintessence could happen without fine-tuning the initial
conditions of the axions and thus explain why the cosmological
constant is nonzero.

While $10^{4-5}$ axions may seem as a large number,
compactifications with such number of axions are known. To assess
whether this many axions could lead to problems, such as the species
problem, we estimate the radiative corrections to the Planck mass
due to the large number of light axions. Requiring that the tree
level contribution to Planck mass dominates the loop contributions
leads to an upper bound on the number of axions and hence an upper
bound on the number of Hubble times of cosmic acceleration. We
perform this estimate and find that the axion quintessence is well
in the range allowed by the species problem.

Many axions have been previously discussed in the related problem of
achieving inflation in string theory in \DimopoulosAC.\ The many
axion quintessence as an effective field theory was studied in
\KaloperAJ.\ In a sense, some of the focus of the present paper is
to determine whether this effective field theory  works when
embedded into string theory. The physics of many axion quintessence
and inflation is the same, although the two mechanisms happen in a
different regions of axion parameter space. Hence, it could be that
some of the string theory axions have driven inflation while others
are currently responsible for cosmological constant. For recent
proposals of other dynamical mechanisms to solve the cosmological
constant problem, see \refs{\SteinhardtBF,\ItzhakiRE}.

This paper is organized as follows. In section 2, we discuss the
generation of axion potential in string theory and the conditions
under which this potential is comparable to the cosmological
constant. In section 3, which is the main part of the paper, we
discuss the cosmology of the axion quintessence, determining the
conditions under which it can be realized in string theory. In this
section we borrow results for string theory axions that are derived
for various string compactifications in sections 4-7. In these
sections we also estimate the height of the axion potential and give
an upper bound on the duration of axion quintessence from
consideration of radiative corrections.

\newsec{Very Light Axions in String Theory}

In string theory there are fields with naturally very flat
potential. These are pseudoscalar fields, called axions, that have
the shift symmetry $a\rightarrow a+c$. If this symmetry was exact,
it would set the potential to be independent of $a$ and render the
axion massless. Effects that break this symmetry generate potential
for $a$. In string theory, the shift symmetry is exact to all orders
in perturbation theory so the axions receive potential only from
nonperturbative instanton effects. These effects generate potentials
that are exponentially suppressed by the instanton action. Hence, if
the instantons have large actions, they can give rise to a potential
many orders of magnitude below the Planck scale. In this section, we
will estimate how large the instanton actions have to be to lead to
an axion potential with magnitude around the present value of
cosmological constant.

The instantons break the shift symmetry down to a discrete shift
symmetry $a\rightarrow a+2\pi n$ where $n$ is arbitrary integer.
They generate a superpotential \eqn\axpot{W=M^3 e^{-S_{inst}+ia},}
where $S_{inst}$ is the instanton action and $M$ is the scale of the
instanton physics which could be around the Planck scale or the
string scale.  One can estimate the axion potential by substituting
\axpot\ into the formula for the potential of low-energy effective
$\CN=1$ supergravity \eqn\pots{V=e^{K\over
M_P^2}\left(K^{i\bar{j}}D_iW D_{\bar{j}} \overline{W}-{3\over
M_P^2}|W|^2\right),} where $K$ is the Kahler potential and $W$ is
the superpotential.

If we do not assume low energy supersymmetry breaking, then we
estimate \eqn\vest{V\sim M^4 e^{-S_{inst}}(1-\cos(a))+V_0,} where
$V_0$ represents other contributions to vacuum energy. Since, we
assume that axions are the only very light scalars, all the other
fields will have rolled down to their vacuum well before the dark
energy has started to dominate the density of the universe. It
follows that they contribute only a constant term to the potential,
given by the sum of their vacuum energies. Hence, for our purposes
$V_0$ is a constant.

If we take $M\sim M_P,$ the axion contribution to the vacuum energy
is comparable to the present value of dark energy
\eqn\darke{\rho_{vac}\sim\ e^{-280}M_P^4,} if $S_{inst}\sim 280$.

Low energy supersymmetry breaking can further suppress the axion
potential. To estimate the axion potential, we add to the
superpotential a term $W_0$ from the supersymmetry breaking sector
\eqn\superg{W=M^3e^{-S_{inst}+ia}+W_0.} The axion potential gets a
contribution $V\sim M^2\partial_i W_0e^{-S_{inst}+ia}+ c.c.$ from
interference between the one-instanton term and the supersymmetry
breaking term $\partial_i W\sim m_S^2$. Here $m_S$ is the scale of
supersymmetry breaking. Hence, the axion potential is
\eqn\apotc{V\sim m_S^2M^2e^{-S_{inst}}(1-\cos(a))+V_0.} If
supersymmetry is broken at low energy $m_S\sim 1{\rm TeV},$ we get
$V\sim e^{-280}M_P^4$ for instantons that have  actions
$S_{inst}\sim200.$ In summary contribution of one axion to vacuum
energy is comparable to present dark energy density if the
instantons breaking the shift symmetry have actions in the range
$S_{inst}\sim 200-280.$

With such such large instanton actions, the instanton effects that
would stabilize the moduli that are scalar partners of the axions
are negligible,  so other effects are needed to stabilize them.
These could be for example perturbative effects that leave the
axions massless. A concrete example that uses tree level potential
from RR and NS-NS fluxes to stabilize the moduli has been recently
discussed by Kachru et al. \DeWolfeUU\ in the context of type IIA
string.  We should also point out that supersymmetry breaking is
expected to give the moduli heavy masses. If the SUSY breaking mass
of the moduli comes from perturbative effects that do not break the
shift symmetry, the axions remain massless. Hence the stabilization
of these moduli could be tied to supersymmetry breaking. For
sufficiently large supersymmetry breaking scale, the moduli are
heavy enough to evade various constrains coming from dark matter
density, fifth force and other experiments. Some models where moduli
are stabilized along these lines were recently discussed in
\refs{\ConlonKI,\ConlonTQ}.

\newsec{Axions and the Vacuum Energy}

As discussed in the introduction, observational constraints favor a
pseudoscalar axion quintessence that is protected by shift symmetry.
We will now study the cosmology of such axions, with emphasis on the
conditions that lead to quintessence.

We describe the universe in the flat FRW metric \eqn\frwmetric{ds^2=
-dt^2+ R(t)^2(dx^2+dy^2+dz^2),} where $R(t)$ is the expansion factor
of the universe. The axion is a four-dimensional pseudoscalar field
that is periodic with period $2\pi.$  We assume that it has an
instanton generated potential
\eqn\apot{V(a)=\mu^4(1-\cos({a}))+V_0,} where $\mu$ parametrizes the
scale of the axion potential and $V_0$ represents other
contributions to vacuum energy. As discussed below eq. \vest,\ we
take $V_0$ to be a constant. The action of the axion in flat FRW
universe described with the metric \frwmetric\ is \eqn\aact{S=\int
d^4x R^3\left(-{F_a^2\over 2}
\partial_\mu a\partial^\mu a- \mu^4(1-\cos(a))-V_0\right),} where $F_a$
is the axion decay constant. We note that \aact\ has nonstandard
normalization of the kinetic term. Rescaling the axion $a\rightarrow
a/F_a$ restores canonical normalization and rescales the period of
the axion to $2\pi F_a.$

We assume that the scalar field is homogeneous in space so that its
expectation value depends only on time $a=a(t).$ Varying the action
\aact\ with respect to $a$ gives the equation of motion of the axion
\eqn\emot{\ddot{a}+3H\dot{a}+{\mu^4\over F_a^2}\sin(a)=0,} where
$H=\dot{R}/R$ is the Hubble parameter. In a universe whose density
is dominated by dark energy \eqn\hubbld{H=\sqrt{\rho_{vac}\over 3
M_P^2}.}

The equation \emot\ describes a particle moving in  one-dimensional
potential $V(a)$ with friction force $-3H\dot{a}.$ If the initial
conditions of the axions are generic, the situation is easy to
summarize. For $H\lesssim m_a$, where $m_a^2 = V''/F_a^2=
\mu^4/F_a^2$ is the mass of the axion particles, the axion is
under-damped. The axion oscillates around the minimum of its
potential. These oscillations describe a Bose-Einstein condensate of
axion particles at zero momentum with mass $m_a=\mu^2/F_a.$ In this
case, the axions contribute to cold dark matter
\refs{\PreskillCY,\bb,\cc}.  For $H\gtrsim m_a,$ the axion is
overdamped. Due to the Hubble friction, the axion is slowly rolling
down its potential and contributes to dark energy density. Hence,
the condition for an axion with generic initial conditions to
contribute to dark energy instead of dark matter is \eqn\cona{
H\gtrsim {\mu^2\over F_a}.} The axion contribution to dark energy is
at most $\mu^4$. With the help of \hubbld\ and \cona,\ this becomes
\eqn\avam{\delta\rho_{vac}\lesssim \mu^4\lesssim \rho_{vac}{
F^2_a\over M_P^2}.} The most important point to notice about \avam\
is that the relative contribution of the axion to dark energy
depends only on the ratio of the $F_a$ and $M_P.$ In particular, a
single axion that starts with generic initial conditions can account
for all of dark energy only if it has Planck size axion decay
constant. Hence, to find out whether an axion can account for all of
dark energy, it is crucial to understand what are the possible
values of $F_a$ in string theory.

As we will demonstrate below in various string compactifications,
there is an upper bound on the axion decay constant
\eqn\upperb{F_a\lesssim {x M_P \over S_{inst}},} where $x$ is of
order one and $S_{inst}$ is the action of the instantons that break
the axionic shift symmetry and generate axion potential. It is
interesting to notice that this bound depends only on $M_P$ and
$S_{inst}.$ All the dependence on $g_s, \ell_s$ and the type of
string theory enters the formula only through $M_P, S_{inst}.$ This
simplicity  suggests that the formula \upperb\ is  true in any
four-dimensional string compactification\foot{This bound was
conjectured in \ArkaniHamedDZ\ to hold in any quantum theory of
gravity while this work was in gestation.}.

For moderately large $S_{inst}$, \avam\ together with \upperb\
implies a strong bound on the contribution of one axion to the dark
energy \eqn\acont{\delta\rho_{vac}\lesssim {\rho_{vac}\over
S_{inst}^2}.} For $S_{inst}\sim 200-280$ this gives
$\delta\rho_{vac}\lesssim \rho_{vac}/10^{4-5}.$   This bound
originates from the fact that only axions whose potential is about
$10^{4-5}$ times smaller than the current vacuum energy are lighter
than the Hubble scale and hence contribute to dark energy. These
axions have instanton actions in the range \eqn\instanc{S_{inst}\sim
210-290.} This is slightly larger than the actions we estimated in
previous section because of the extra suppression of the axion
potential by a factor of $10^{4-5}$ compared to the cosmological
constant. Axions with larger potential are under-damped so they
oscillate around the minimum of their potential. They contribute to
dark matter density instead. These axions could potentially generate
too much dark matter, thus over-closing the universe.

Hence, a single axion with generic initial conditions can account
only for a fraction of the present dark energy density.  However,
string compactifications can have many axions in their
four-dimensional spectrum. Each of these axions, if it satisfies the
condition \cona,\ gives an additive contribution to the vacuum
energy. For simplicity, let us assume that all axions have the same
axion decay constant $F_a\sim M_P/S_{inst}.$ If there are $N$
axions, their contribution to vacuum energy adds up to
\eqn\aconp{\delta\rho_{vac, N}\sim \rho_{vac}{\sum_i^N F_{a,i}^2
\over M_P^2} \lesssim {N \rho_{vac}\over S_{inst}^2}.} So in
compactifications that have \eqn\nbond{N\gtrsim S_{inst}^2} axions,
the cosmological constant could be entirely due to axion potential
energy. In string compactifications we expect the instanton actions
$S_{inst}$ of different instantons to vary by a factor of order one.
Hence only some of them will fall into the preferred range \instanc\
and the actual number of axions necessary for quintessence is
somewhat larger than the estimate \nbond.\ For an extensive
discussion of this issue in the related context of axion inflation,
see \EastherZR.\

Known string theory Calabi-Yau three-fold compactifications have up
to $\sim 10^{3-4}$ axions in their four-dimensional spectrum. There
are examples of F-theory compactified on Calabi-Yau four-folds that
lead to $\sim 10^{5-6}$ axions \nref\douglas{M. Douglas, Private
Communication.} \douglas.\ Hence, the quintessence with many axions
could explain cosmological constant in some string
compactifications.

If we assume that the entire vacuum energy is due to the axions, so
that $V_0=0$ in \apot,\  the axion quintessence lasts only for a
finite number of Hubble times. As an illustration of this, let us
assume that all axions have the same potential \apot.\ A simple
calculation shows that the number of e-foldings grows as
\refs{\DimopoulosAC,\EastherZR} \eqn\efold{N_{e-fold}\sim {\sum_i^N
F_{a,i}^2\over M_P^2}\lesssim {N\over S_{inst}^2,}} where we used
the estimate \upperb\ for $F_{a,i}.$ So in compactifications with
large number of axions, the axion quintessence could last for
several Hubble times.

It is expected that  $F_a$'s  of different axions are not equal but
rather differ by factors of order one. Let us discuss how this
variation affects \efold.\ The number of e-foldings of quintessence
due to axions \efold\ depends on the average of the squares of axion
decay constants, hence it is insensitive to the individual variation
of $F_a.$ We expect that the estimate \upperb\ gives the average
$F_a$ correct up to a factor of order one. Hence we can trust the
estimate \efold\ for the number of e-foldings up to a factor or
order one.

In string compactifications with a small number of axions, \nbond\
leads us to conclude that only a small fraction of vacuum energy is
due to axions, if the axion initial conditions are generic. In these
compactifications, the cosmological constant is due to the constant
piece $V_0$ of vacuum energy \vest\ that represents the energy of
the ``true vacuum'' in which all scalars have already reached the
minima of their potentials.

To close this section, let us return back to the heavier axions.
Under generic initial conditions these axions contribute to dark
matter. However, for special initial conditions the axions could
contribute to dark energy density. If their initial conditions are
fine-tuned so that the axions stay at the top of their potential for
at least one Hubble time, they will behave like dark energy instead.
Let us estimate the necessary fine-tuning. For simplicity, we assume
that there is only one such heavy axion. The axion is slow-rolling
on the flat portion of its potential near the top so that it
contributes to dark energy. The equation of motion for the deviation
$\delta a=\pi-a$ of the axion from the top of its potential  is
\emot\ \eqn\ripto{3H \dot{a}+{\mu^4\over F_a^2} \delta a=0.} The
axion stays close to the top, $|\delta a|\lesssim 1$ for at least
one Hubble time $t_H\sim 1/H$ if the axion speed is less than the
Hubble scale $\dot{a}\lesssim H.$ Combining this with \hubbld\ and
\ripto\ gives a bound on the initial axion displacement
\eqn\achto{\delta a_{init}\lesssim {F^2_a\over
M_P^2}{\rho_{vac}\over \mu^4}\sim {1\over S_{inst}^2}
{\rho_{vac}\over \mu^4}.} If we assume that the entire cosmological
constant is due to potential energy of a single axion then
$\rho_{vac}\sim \mu^4$  so the axion initial conditions have to be
fine-tuned to one part in $S_{inst}^2\sim 10^{4-5}$
\eqn\bachto{\delta a_{init} \lesssim {1\over S_{inst}^2}.} In
\KaloperAJ\ it has been argued that the quantum fluctuations could
perturb the axion away from the maximum of the potential, thus
spoiling the fine-tuning of the initial conditions.

In summary, the cosmological constant could be due to axion
quintessence if there are at least $S_{inst}^2\sim 10^{4-5}$ axions.
In compactifications with fewer axions, the cosmological constant
could still be due to axions, modulo issues with quantum
fluctuations, if the initial conditions of say one of them are
fine-tuned to one part in $S_{inst}^2\sim 10^{4-5}$.

In the following four sections, we discuss the parameters of axions
in different string compactifications. Among other results, we
derive the upper bound \upperb\ on the axion decay constant in these
compactifications.

\newsec{Heterotic String Theory}

\lref\thomas{P. Fox, A. Pierce, and S. Thomas, ``Probing A QCD
String Axion With Precision Cosmological Measurements,''
hep-th/0409059.}

In heterotic string theory, the axions come from zero modes of the
NS-NS $B$-field. The components of $B$-field polarized along the
four noncompact dimensions are dual to a pseudoscalar conventionally
called the model independent axion. The model independent axion has
axion decay constant \refs{\thomas ,\axionpaper} $F_a=\alpha_G
M_P/2\sqrt{2}\pi,$  where $\alpha_G\sim 1/25$ is the unified gauge
coupling. We recall that in four-dimensional compactifications of
heterotic string theory on a six-manifold $X$ the gauge coupling and
the Planck scale are \eqn\recam{ M_P^2={4\pi V_X\over g_s^2
\ell_s^8} \qquad \alpha_G={g_s^2\ell_s^6\over V_X}.}  This can be
deduced by dimensional reduction of the ten-dimensional gauge action
$-{1/ 4(2\pi)g_s^2 \ell_s^6}\int \tr F\wedge \star F$ and the
gravity action ${2\pi/g_s^2\ell_s^8}\int d^{10}x \sqrt{-g} R$. The
instantons that break the shift symmetry of the model-independent
axion are NS5-branes wrapped around $X.$ Their action is
$S_{inst}=2\pi V_X/g_s^2=2\pi/\alpha_G.$ For $\alpha_G\sim1/25$ this
gives $S_{inst}\sim157.$ In terms of $S_{inst}$, the axion decay
constant becomes \eqn\dacmp{F_a={M_P\over \sqrt{2} S_{inst}}} in
agreement with the general conjecture \upperb.\ Numerically, \dacmp\
gives $F_a\sim 1.1\times 10^{16}\GeV.$ The axion decay constant is
two orders of magnitude below $M_P$ so the model independent axion
alone does not lead to quintessence unless the initial conditions of
the axion are fine-tuned.

Hence, we turn our attention to model dependent axions. The model
dependent axions come from zero modes of the NS-NS $B$-field
polarized within the compactification manifold. If $\omega_i,
i=1,\dots,b_2(X)$ are harmonic two-forms normalized so that
\eqn\normap{\int_{C_i} \omega_j=\delta_{ij},} where $C_i$ is a basis
of $H_2(X,\Bbb{Z})$ modulo torsion, then the axions are the
four-dimensional fields $a_i$ in the ansatz \eqn\anbas{B=\sum_i a_i
{\omega_i\over 2\pi}.} Dimensional reduction of the kinetic term of
the $B$-field $-{2\pi\over g_s^2\ell_s^4}\int H\wedge \star H$ to
four dimensions gives the kinetic energy of the axions
$S_{kin}=-\half \sum_{i,j}\int d^4x  \gamma_{ij}
\partial_\mu a_i\partial^\mu a_j,$ where
\eqn\andale{\gamma_{ij}={1\over 2\pi g_s^2\ell_s^4}\int_X
\omega_i\wedge \star\omega_j.} Let us estimate the axion decay
constant of a generic model dependent axion in heterotic string
theory. The axion comes from a zero mode of a harmonic two-from
$\beta$ that is dual to a two-cycle $C,$ so $\int_C \omega=1.$ If
$R=V_C^{1/2}$ is the size of $C$, then the integral $\int_X
\omega_i\wedge \star\omega_j$ scales as $R^2$ so we estimate the
axion decay constant to be \eqn\fgehe{F_a={x R\over \sqrt{2\pi}
g_s\ell_s^2},} where $x$ is a dimensionless model dependent constant
of order one. The instantons that violate the shift symmetry
$a\rightarrow a+c,$ are worldsheet instantons wrapping $C$. The
action of these instantons is \eqn\inpac{S_{inst}={2\pi R^2\over
\ell_s^2}.}  Using this we express the axion decay constant in terms
of $S_{inst}$ and $M_P$ \eqn\expresm{ F_a={xM_P\over S_{inst}}
\sqrt{ R^6\over 2V_X}\lesssim {x M_P\over \sqrt{2} S_{inst}}.} The
last inequality comes from $R^6\lesssim V_X,$ since the size of the
curve $C$ cannot be greater than the size $R_X\sim V_X^{1/6}$ of the
compactification manifold $X.$

\medskip\noindent{\it Constraints from Particle Physics}

Heterotic string theory leads naturally to four-dimensional models
of particle physics based on unified gauge groups. Here, we will use
the additional constraints coming from matching to four-dimensional
gauge couplings to see that axions quintessence does not work in
weakly coupled heterotic string theory. Similar argument has been
advanced in \refs{\ChoiXN}.

For heterotic string, the volume of the compactification manifold is
determined in terms of the gauge coupling and string parameters
\recam\ \eqn\ropac{V_X={g_s^2\ell_s^6\over \alpha_G}= 25 g_s^2
\ell_s^6.} The weakly coupled perturbative description is valid for
$g_s\lesssim 1$ which gives an upper bound on the volume of the
compactification manifold $V_X\lesssim 25 \ell_s^6.$ This has severe
implications for the flatness of the axion potential. The worldsheet
instantons breaking the axionic shift symmetry have actions that are
proportional to the area of the curve they wrap. For small $V_X,$
the areas of the curves are small and the axion potential due to
worldsheet instantons is not sufficiently suppressed. For a generic
axion the instanton action \inpac\ is bounded above because
$R\lesssim V_X^{1/6}$, so \eqn\axti{S_{inst}\lesssim{2\pi
V_X^{1/3}\over\ell_{s}^2}={2\pi g_s^{2/3}\over \alpha_G^{1/3}}=18
g_s^{2/3},} where we used \ropac\ to reexpress $V_X$ in terms of
$\alpha_G$. In weakly coupled heterotic string we take $g_s\lesssim
1$ which gives\foot{Larger values of $S_{inst}$ are possible for
some axions for example in anisotropic CY manifolds \axionpaper.\
However, here we are interested in generic axions because, as
discussed in section 3, a large number of axions is necessary for
quintessence.} $S_{inst}\lesssim 18.$ The worldsheet instanton
actions are one order of magnitude below the range $210\sim 290$
necessary for getting sufficiently flat axion potential for
quintessence. Hence the axions get large instanton generated
potential and are not suitable for explaining present day cosmic
acceleration.

\newsec{Heterotic M-theory}

In weakly coupled heterotic string theory we found that the
instanton actions \axti\ of instantons that generate axion potential
are too small. In strongly coupled heterotic string, one can achieve
larger instanton actions \refs{\banksone, \ChoiXN}, since the
instanton action \axti\ grows as a positive power of $g_s.$ When the
string coupling gets large, the heterotic string is better described
using a dual description as M-theory compactified on $X\times I$
where $X$ is the compactification manifold of the heterotic string
and $I$ is an interval.  The length of the interval grows with the
string coupling. The $E_8\times E_8$ gauge symmetry lives on the
boundaries of $I$. The world-sheet instantons of heterotic string
become open M2-branes stretched across the interval. Their action is
proportional to the length of the interval. Hence, for sufficiently
long $I,$ these actions are large and the axion potential can be
exponentially suppressed down to current dark energy density. For
simplicity we will consider the case when $X\times I$ has the
product metric. In more general heterotic M-theory
compactifications, the metric along $I$ is warped. We do not expect
the warping to significantly affect our results.

Let us discuss some features of these compactifications. For further
details on axions in heterotic M-theory and discussion of the
conventions, see \axionpaper.\ We define the M-theory length
$\ell_{11}=(4\pi\kappa^2)^{1/9}$ and mass $M_{11}=\ell_{11}^{-1}$.
We normalize $G$-form field  so that it has integer periods. With
these conventions, the action of eleven-dimensional supergravity
becomes \eqn\elema{S_{11}=2\pi \int\left({1\over \ell_{11}^9}
d^{11}x\sqrt{-g} R-{1\over 2\ell_{11}^3} G\wedge \star G-{1\over 6}
C\wedge G\wedge G\right).} The gauge fields live on the boundary of
the interval. Their action is \eqn\hetga{S_{YM}=-{1\over 4(2\pi)
\ell_{11}^6}\int \tr F\wedge \star F.} Dimensional reduction of
\elema\ and \hetga\ leads to four-dimensional gauge coupling and
Planck mass \eqn\gapa{ \alpha_G={\ell_{11}^6\over V_X},\qquad
M_P^2={4\pi V_X L\over \ell_{11}^9},} where $L$ is the length of the
interval $I.$ Note that in eleven-dimensional Planck units, the
volume of the Calabi-Yau manifold $X$ is determined by the unified
gauge coupling $V_X=\alpha_G^{-1}\ell_{11}^6\sim 25 \ell_{11}^6.$

In heterotic M-theory, there is a model independent axion and
several model dependent axions. As with weakly coupled heterotic
string, we concentrate on the model dependent axions. These receive
potential from membrane instantons stretched between the two
boundaries and wrapping a curve in the Calabi-Yau manifold. The
number and the properties of model dependent axions depends on the
compactification. The model dependent axions come from modes of the
three-form field with one index along $I$ and two along  $X$
\eqn\hemo{ C= \sum_i a_i \omega_i {dx^{11}\over 2\pi L}.} For an
axion coming from a generic curve $C$ with $\int_C \omega =1,$
dimensional reduction of the $G$-field kinetic energy \elema\ gives
the axion decay constant \eqn\genap{F_a={M_P\over 2\sqrt{2}\pi}
{\ell_{11}^3\over V_C L}={M_P\over \sqrt{2} S_{inst}},} where $V_C$
is the volume of the curve $C$ and $L$ is the length of the M-theory
interval. $S_{inst}=2\pi V_C L/\ell_{11}^3$ is the action of the
M2-brane instanton wrapping $C\times I.$ These instanton breaks the
shift symmetry of the axion to $a\rightarrow a+2\pi$ and give the
axion a nonzero mass.

\subsec{\it Unification and the Cosmological Constant}

Compactifications of heterotic M-theory naturally incorporate
unification of gauge couplings. Matching to the experimental values
of the unified gauge coupling and the unification scale determines
some parameters of the compactification. This allows us to estimate
the scale of the axion potential by estimating the instanton actions
of the membrane instantons that break the axionic shift symmetry.

Recall that in heterotic M-theory, the gauge fields live on the
boundary of $I.$ The unified gauge group is embedded into one of the
two boundary $E_8$ gauge theories. In the usual approach to
phenomenology, the unified gauge group is broken to  standard model
gauge group using discrete Wilson lines. Hence, the unification
scale is set by inverse radius of the Wilson lines which is roughly
the inverse radius of $X$ \eqn\mguta{M_{GUT}\sim {1\over R_X}\sim
{\alpha_G^{1/6}\over \ell_{11}}.}

Consider generic axion corresponding to $C\times I$ where $C$ is a
curve in $X$. For a generic curve $C$, $V_C\sim
V_X^{1/3}=\alpha_G^{-1/3} \ell_{11}^2,$ so the action of an M2-brane
wrapping $C\times I$ is \eqn\agen{S_{inst}={2\pi V_C L\over
\ell_{11}^3}\sim{2\pi\over \alpha_G^{1/3}} {L\over \ell_{11}}.} This
becomes, in terms of the unification scale \mguta\ and the Planck
scale \gapa,\ \eqn\instach{S_{inst}\sim {\alpha_G\over 2}
{M_P^2\over M_{GUT}^2}.} If we take for the unification scale
$M_{GUT}=2\times 10^{16}\GeV,$ we find that the instantons have
actions around $S_{inst}\sim 290,$ which is at the upper end of the
range $S_{inst}\sim 210-290$ preferred for axion quintessence. If
the metric is warped along the interval $I$, the instanton actions
can be somewhat different. As discussed below \nbond,\ in generic
compactifications we expect the instanton actions $S_{inst}$ of
different membrane instantons to vary by factor of order one. Hence
only some of them will fall into the range $210\sim 290$ and the
number of axions necessary for quintessence is somewhat larger than
the estimate $10^{4-5}$ \nbond.\ This remark applies also to axions
in other string compactifications that are considered in the
following sections.

Hence in the context of axion quintessence in heterotic M-theory,
the hierarchy between the cosmological constant and the Planck scale
is related to the hierarchy between the GUT scale and the Planck
scale. The estimated value of the cosmological constant can be close
to the observed value. Whether that is the case, depends on the
details of supersymmetry breaking and the generation of the axion
potential that were discussed in section 2.

One can derive similar estimates for cosmological constant and the
unification scale in other string compactifications like $G_2$
holonomy and type II string compactifications that we study in the
following sections. In these compactifications, the cosmological
constant gets again related to the hierarchy between the unification
scale and the Planck scale. Intuitively, this is because the
hierarchy between $M_{GUT}$ and $M_P$ comes from increasing the size
of the compactification manifold which increases the instanton
actions, since these are given by volumes of various submanifolds of
$X$.

\subsec{Axion Quintessence and the Species Problem}

$10^{4-5}$ axions may seem quite a large number of light scalars.
The light axions radiatively induce corrections to the Planck scale
\eqn\plancc{\delta M_P^2\simeq \pm {N\Lambda_{UV}^2\over 16\pi^2},}
where $N$ is the number of axions and $\Lambda_{UV}$ is a high
energy cut-off scale. For large number of axions $N$, this could be
comparable to the tree level value of Planck scale \gapa.\ A good
measure of whether we can trust our tree-level estimates of axion
parameters is whether the radiative corrections \plancc\  to the
Planck scale are suppressed compared to the tree-level contribution.
Requiring that these corrections are suppressed leads to an upper
bound on the number of axions and hence an upper bound on the
duration of quintessence.

In the context of the heterotic M-theory, we take the high-energy
cut-off scale to be the M-theory scale $\Lambda_{UV}\sim
1/\ell_{11}.$ Requiring that the radiative corrections are smaller
than the tree-level expression for Planck mass gives $N\lesssim
16\pi^2( M_P \ell_{11})^2,$ which with the help of \gapa\ and \agen\
becomes \eqn\nhebo{N\lesssim {32\pi^2\over \alpha_G^{2/3}}
S_{inst}.} This leads to an upper bound on the duration of the axion
quintessence \efold\ \eqn\dubao{N_{e-fold}\lesssim {32\pi^2\over
\alpha_G^{2/3} S_{inst}}.} One thing to notice about \dubao\ is that
the number of Hubble times of accelerated expansions decreases with
$S_{inst}$ and hence with the energy of the axion potential. Similar
formulas are valid in other string compactifications considered in
the following sections. For instanton actions $S_{inst}\sim
210-290,$ this gives $N_{e-fold}\lesssim 10-13.$ Hence quintessence
could last for a few Hubble times in heterotic M-theory.

\bigskip\noindent{\it Axion Inflation and the Species Problem}

The bound \dubao\ also gives a limit on the number of Hubble times
of inflation in the axion model of inflation of Dimopoulos et al.
\DimopoulosAC.\ In that model, the axion potential is used to
explain the inflation in the early universe. If we assume that the
inflationary potential has height around the GUT scale $V\sim
(10^{16}\GeV)^4,$ then the actions of instantons that generate this
potential are roughly $S_{inst}\sim \ln(M_P^4/V)\sim 20.$ Here
\dubao\ gives an upper bound on the duration of axion inflation
$N_{e-fold}\lesssim 130$ which is above the minimum of $ 60$ e-folds
required by present day cosmological experiments.

\newsec{M-Theory on $G_2$ Holonomy Manifolds}

We assume that M-theory is compactified on a manifold $X$ of $G_2$
holonomy with volume $V_X.$ The four-dimensional effective theory
has ${\cal N}=1$ supersymmetry and can lead to semi-realistic models
of particle physics \refs{\AcharyaGY,\FriedmannTY}.

The axions come from zero modes of the three-form field $C$. If
$D_i,i=1,\dots b_3(X)$ is a basis of three-cycles of $X$ and
$\omega_i, i=1,\dots, b_3(X)$ is the dual basis of harmonic
three-forms on $X$  such that $\int_{D_i} \omega_j=\delta_{ij},$
then the axions $a_i$ are four-dimensional fields coming from the
ansatz \eqn\anap{C={1\over 2\pi} \sum_i a_i \omega_i.} The kinetic
energy of the axions comes from  dimensional reduction of \elema\
\eqn\kinap{S_{kin}=-{1\over 2\pi \ell_{11}^3}\int d^4x \half
\partial_\mu a_i \partial^\mu a_j \int \omega_i\wedge
\star\omega_j.} The generic axion $a$ has axion decay constant
roughly \eqn\gtwog{F_a^2={x^2\over 2\pi}{R\over \ell_{11}^3},} where
$R=V_{D}^{1/3}$ is the size of the three-cycle $D$ and $x$ is a
dimensionless model dependent number of order one.

The axion gets potential from M2-brane instanton wrapping
three-cycle $D$. The membrane instanton has action
\eqn\inag{S_{inst}=2\pi {R^3\over \ell_{11}^3}.} For quintessence,
we need $S_{inst}\sim210-290,$ which determines the size of the
three-cycles \eqn\rar{R=\ell_{11}\left(S_{inst}\over
2\pi\right)^{1/3},} or numerically $R\sim 3 \ell_{11}$. Combining
\gtwog\ and \rar\ we express the axion decay constant in terms of
$M_P$ and $S_{inst}$ \eqn\asta{F_a={x M_P\over \sqrt{2}
S_{inst}}\sqrt{R^7\over V_X}\lesssim {xM_P\over \sqrt{2} S_{inst}},}
since  $R\lesssim V_X^{1/7}$ as the size of the cycle $D$ is bounded
by the size of $X.$ In \asta\ we used that the Planck scale is
$M_P^2=4\pi V_X/\ell_{11}^9$ which follows from dimensional
reduction of the eleven-dimensional supergravity action \elema.\

\subsec{Constraints from Unification}

$G_2$ holonomy compactifications naturally implement models of
particle physics based on unified gauge groups
\refs{\AcharyaGY,\FriedmannTY}. Matching to the observed value of
gauge coupling and unification scale gives us further constraints on
the compactifications  that help estimate the size of the instanton
actions. This gives a measure of what is the likely value of axion
potential in $G_2$ holonomy compactifications.

In these models, gauge symmetry comes from an orbifold singularity
that is fibered along a three-cycle $Q$ in $X$. The unified gauge
coupling is \FriedmannTY\ \eqn\uniga{\alpha_G={\ell_{11}^3\over
V_Q}.} If we assume that the unified gauge group is broken down to
the standard model gauge group using discrete Wilson lines, the
unification scale is given rougly by the inverse radius of $Q$.
After taking into account threshold corrections, the relation
between the GUT scale and the volume $V_Q$ is \FriedmannTY\
\eqn\gata{M_{GUT}^3={L(Q)\over V_Q},} where $L(Q)$ is Reidemeister
or Ray-Singer torsion, which is a topological invariant that depends
on $Q$ and on the Wilson line on $Q$ that breaks $SU(5)$ down to the
standard model gauge group. For example $Q$ could be a lens space
$S^3/Z_q$, with Wilson line on $Q$ that has eigenvalues $\exp(2\pi i
\Delta_i/q)$ with $\Delta_i=(2w,2w,2w,-3w,-3w),$ with $w,q$ coprime
integers, $L(Q)=4q\sin^2(5\pi w/q).$ For the minimal choice
$q=2,w=1,$ one has $L(Q)=8.$  Combining \uniga\ and \gata,\ we find
the M-theory scale in terms of low-energy parameters
\eqn\mpal{M_{11}={M_{GUT}\over (\alpha_G L(Q))^{1/3}}.}

A generic three-cycle $D$ has volume roughly $V_D\sim V_X^{3/7}$.
The action of a  membrane instanton that wraps $D$ is $S_{inst}=2\pi
V_D/\ell_{11}^3.$ In terms of the Planck scale and the unification
scale this is \eqn\agam{S_{inst}\sim 2\pi \left((\alpha_G
L(Q))^{2/3} M_P^2\over 4\pi M_{GUT}^2\right)^{3/7}} or numerically
$S_{inst}\sim 90.$ This is two to three times below the preferred
range $S_{inst}\sim 210-290.$

\newsec{Type II Compactifications}

Let us briefly review the calculation of the decay constant of the
axions in  type II string compactifications \axionpaper.\ The axions
come from dimensional reduction of $q$-form RR-field $C_q$. The
axions are the four-dimensional fields $a_i$ in the ansatz
\eqn\ansam{C_q=\sum_i {a_i\over 2\pi} \omega_i, \quad i=1,\dots,
b_q(X),} where $\omega_i$ are harmonic forms normalized so that
$\int_{C_i} \omega_j=\delta_{ij},$ where $C_i$ is a basis of
$H_q(X).$ The kinetic energy of axions comes from dimensional
reduction of the kinetic term of the $q$-form field
\eqn\amar{-{2\pi\over \ell_s^{8-2q}}\int_{R^4\times X} F_{q+1}\wedge
\star F_{q+1}.} Substituting \ansam\ into \amar\ one finds
$S_{kin}=-\half \sum_{i,j} \int d^4x \gamma_{ij}
\partial_\mu a_i \partial^\mu a_j,$ where \eqn\gamap{\gamma_{ij}={1\over 2\pi \ell_s^{8-2q}}\int_X
\omega_i\wedge \star \omega_j.} If the axion comes from cycle $C$,
the integral $\int_X \omega\wedge \star \omega$ scales as
$R^{6-2q},$ where $R= V_C^{1/q}$ is the size of $C$. Hence the axion
decay constant is roughly \eqn\astra{F^2_a={x^2\over 2\pi}{
R^{6-2q}\over \ell_s^{8-2q}},} where $x$ is a dimensionless number
of order one. The axionic shift symmetry $a\rightarrow a+c$ is
preserved to all orders in string perturbation theory.
Nonperturbatively, the shift symmetry gets broken to a discrete
symmetry $a\rightarrow a+2\pi$ by Euclidean $D(q-1)$ instantons that
wrap $C$. The action of the D(q-1)-brane instanton that generates
the axion potential is \eqn\instam{S_{inst}={2\pi\over g_s}
\left({R\over\ell_{s}}\right)^{q}.} One gets instanton actions in
the preferred range $S_{inst}\sim210- 290$ if the size of $C$ is
roughly a few string lengths \eqn\sizei{R=\ell_s \left(g_s
S_{inst}\over 2\pi\right)^{1/q}.}

The four-dimensional Planck scale $M_P^2=4\pi V_X/g_s^2 \ell_s^8$
follows by dimensional reduction of the gravity action ${2\pi\over
g_s^2\ell_s^8}\int d^{10}x \sqrt{-g} R.$ In terms of $M_P$, the
axion decay constant is \eqn\fexe{F={xM_P\over
\sqrt{2}S_{inst}}\sqrt{R^6\over V_X}\lesssim {x M_P\over \sqrt{2}
S_{inst}}} since the size of the cycle $C$ is bounded by the size of
the compactification manifold $R\lesssim V_X^{1/6}.$

\subsec{Constraints from Unification}

To assess the scale of the axion potential, we estimate the action
of a generic D-brane instanton that breaks the axion shift symmetry.
Assuming that the string compactification implements unification
leads to a more precise estimate of the parameters of the string
compactification and hence of the instanton actions. The gauge
symmetry in Type II D-brane models lives on a stack of D-branes
wrapping a cycle $Q$ in $X$. The four-dimensional gauge coupling is
$\alpha_G=g_s\ell_s^q/V_Q$. If the gauge symmetry is broken down to
standard model gauge symmetry using discrete Wilson lines, the
unification scale is given roughly by the inverse radius of $Q$
\eqn\uscal{M_{GUT}\sim {1\over V_Q^{1/q}}={\alpha_G^{1/q}\over
g_s^{1/q}\ell_s}.} A D-brane instanton that wraps a generic cycle
$C$ has action $S_{inst}=2\pi V_C/g_s\ell_s^q,$ which for $V_C\sim
V_X^{q/6}$ becomes in terms of the Planck scale and the unification
scale \eqn\sitwo{S_{inst}\sim 2\pi {g_s^{q-4\over3}
\alpha_G^{1/3}\over (4\pi)^{q\over 6}}\left(M_p\over
M_{GUT}\right)^{q/3}.} For type IIA D6-brane models, $q=3$ gives
\eqn\dthre{S_{D2}\sim {\pi^{1/2} \alpha_G^{1/3}\over
g_s^{1/3}}{M_P\over M_{GUT}},} which numerically evaluates to
$S_{inst}\sim 73 g_s^{-1/3}.$ Hence for moderately small string
coupling, the instanton action is in the preferred range
$S_{inst}\sim 210-290.$ For type IIB models with gauge symmetry
coming from D7-branes, we get \eqn\twobe{S_{D3}\sim
\left(\pi\alpha_G\over 2\right)^{1/3}\left(M_P\over
M_{GUT}\right)^{4/3}.} Note that this answer depends only on the
phenomenologically observed parameters. In particular it does not
depend on the string coupling. Numerically, we have $S_{inst}\sim
240$ which is in the range preferred for axion quintessence.

\bigskip

We would like to thank M. Douglas, J. McGreevy, S. Kachru, P.
Steinhardt for useful discussions and correspondence and especially
to E. Witten for suggesting the problem, very useful discussions and
careful reading of the manuscript. This work was supported in part
by NSF Grant PHY-0244728, and the DOE under contract
DE-AC02-76SF00515.

\listrefs
\end